# Single-pass, nonlinear frequency conversion of full Poincaré beams

K. Subith,[1,2,*] Ravi K. Saripalli,[1,3] Anirban Ghosh,[1] G. K. Samanta[1]

[1]*Photonic Sciences Lab., Physical Research Laboratory, Navarangpura, Ahmedabad 380009, Gujarat, India*
[2]*Indian Institute of Technology-Gandhinagar, Ahmedabad 382424, Gujarat, India*
[3]*Current address: SLAC National Accelerator Laboratory, 2575 Sand Hill Rd, Menlo Park, California 94025, USA*
*Corresponding author: subith@prl.res.in

**Full Poincaré (FP) beams, a special class of fully correlated beams generated through the coaxial superposition of a Laguerre-Gauss and fundamental Gaussian modes of orthogonal polarizations, contain all possible polarization states on the surface of the Poincaré sphere in a single beam. While the presence of all unconventional polarization states makes the FP beams useful for various applications, the dependence of the refractive index on the polarization restricts the efficient generation of FP beams across the electromagnetic spectrum through nonlinear frequency processes. To avoid such difficulty, we use two contiguous BIBO crystals with orthogonal optic axes and generate an ultrafast FP beam at 405 nm with average power as high as 18.3 mW at a single-pass conversion efficiency of 2.19%. Using Stokes parameters and Stokes phases, we observed the doubling of C-points and L-lines singularities and the orbital angular momentum in the SHG process. We have also devised a new technique to estimate the polarization coverage of the pump and SHG FP beams and observed the variation in polarization coverage with the intensity weightage of the constituting beams. Interestingly, the SHG beam has the highest polarization coverage for the FP beam of equal intensity weightage of the superposed beams. We validated our experimental results in close agreement with the theoretical results. © 2022 Optical Society of America.**

Light beams with spatially varying polarization are commonly known as vector beams or Poincaré beams. Based on the non-separable superposition of two different degree-of-freedom, such as spin angular momentum and orbital angular momentum, the family of Poincaré beams has attracted a great deal of interest due to their vast range of applications. On the other hand, a special class of fully correlated optical beams, defined as the FP beams, are generated through the coaxial superposition of a Laguerre-Gauss and fundamental Gaussian modes of orthogonal polarizations. These beams carry all possible polarization states spanning over the entire surface of the Poincaré sphere. Since the first demonstration [1], the FP beams have attracted a great deal of interest in various science and technology applications, including beam shaping [2], optical trapping [3], ultra-sensitive polarimetry [4], and free-space optical communication [5]. Although the FP beam, in theory, can carry all possible polarization states on the surface of the Poincaré sphere, in practice, due to the finite beam size of the superposed beams, the generation of a full Poincaré beam with full polarization coverage is difficult. Therefore, for all practical purposes, the beam with polarization coverage of more than 75% of the Poincaré sphere is considered as FP beam [6]. Like all the beams of the Poincaré family, the FP beams are typically generated either directly from stress-engineered optical elements [1], a spatial light modulator (SLM) [2], or by using mode conversion in a polarization-based interferometer [5,6]. However, to avoid the limitations of the low damage threshold and the material dispersion, nonlinear frequency conversion techniques can be employed as the direct route to generate FP beams across the electromagnetic spectrum, especially at shorter wavelengths in all time scales. But, the presence of all possible polarization states in the FP beam and the dependence of the refractive index of the nonlinear medium on the polarization state of the interacting beams remained the critical constraints to overcome for efficient nonlinear conversion of FP beams. Recently, we have devised a new experimental scheme for the single-pass second harmonic generation (SHG) of a vector vortex beam using a dual-crystal scheme [7]. Using the same crystal geometry, we have studied the nonlinear interaction of the FP beam in the single-pass SHG process. Although efforts have been made to study the nonlinear interaction of the FP beam [8], many of the interesting effects of scientific importance, such as the effect of intrinsic polarization coverage of the FP beam in the nonlinear process, the conservation of C-points, and L-lines singularities, and the conversion efficiency remained unexplored.

The Stokes parameters have played a crucial role in visualizing the unconventional polarization states of the light beam while projected on the Poincaré sphere. It is well known that the FP beams from the hypothetical complex fields, $S_{12} = S_1 + iS_2$, $S_{23} = S_2 + iS_3$, and $S_{31} = S_3 + iS_1$ derived from the Stokes parameters, $S_1$ $S_2$, and $S_3$ [9]. The Stokes phases, the argument of the Stokes fields, can be calculated as $\phi_{ij} = \tan^{-1}(S_j/S_i)$. Here, $i, j$ = 1, 2 and 3. The vortices present in the Stokes phases are known as Stokes vortices [9,10]. While the Stokes phase, $\phi_{12}$, represents the C-point singularity (the point across which the orientation of the polarization ellipse is undefined), the Stokes field, $S_{23}$, also known as the Poincaré field, represents the net orbital angular momentum of the beam. The number of C-point singularity of the FP beam is the same as the vortex order of the beam. Using the Stokes parameters, we report, for the first time to the best of our knowledge, on the theoretical and experimental study of the single-pass polarization and frequency-doubling characteristics of the FP beam. We have also devised a new scheme to estimate the polarization coverage of the Poincaré beam of different orders and studied the variation of polarization coverage in the nonlinear process.

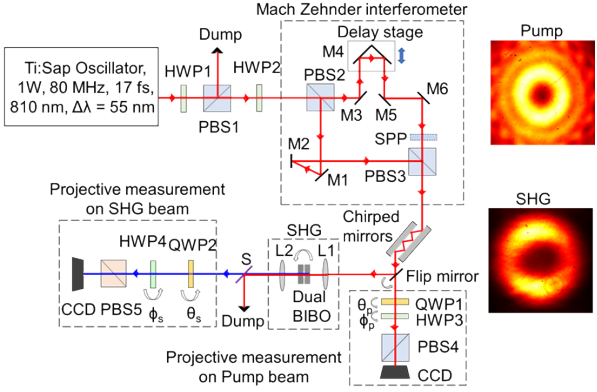

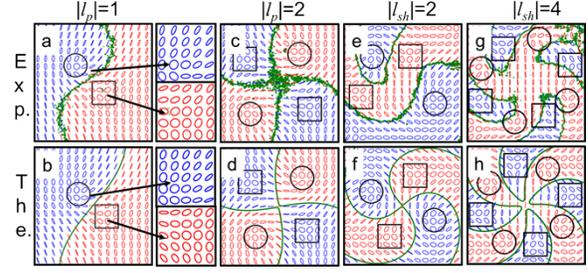

Fig.1: Experimental setup for SHG of full Poincaré beam. HWP1-4: λ/2 plates; PBS1-5: polarizing beam splitter cubes; SPP: spiral phase plate; QWP1-2: λ/4 plates; M1-6: mirrors; L1-2: lenses; S: wavelength separator; CCD: camera; dual-BIBO: nonlinear crystal. Intensity profile of pump and SH FP beams.

Fig. 2: (a) Experimental and (b) theoretical transverse polarization distribution of the FP pump beam of vortex order, $|l_p| = 1$. Second column: the magnified images of the lemon and star (C-point) singularities highlighted by the black circle and square, respectively. (Experimental, theoretical) polarization pattern (c, d) for pump of order, $|l_p| = 2$, (e, f) for SHG of pump order, $|l_p| = 1$, and (g, h) for SHG beam of pump order, $|l_p| = 2$. The green lines are L-lines, and black lines guide the eyes.

The schematic of the experimental setup is shown in Fig. 1. A Ti: Sapphire laser delivering ultrafast pulses of width ~17 fs at a repetition rate of 80 MHz with an average output power of 870 mW is used as the fundamental source. The output radiation has a spectral bandwidth of 55 nm centered at 810 nm. The combination of a λ/2 plate (HWP1) and a polarizing beam splitter (PBS1) cube is used to control laser power in the experiment. The second λ/2 plate (HWP2) is used to manage the polarization of the input beam to the polarization-based Mach-Zehnder interferometer (MZI) comprised of two PBS cubes (PBS2 and PBS3) and a set of plane mirrors, M1-6. A delay stage is used to match the path lengths of the two arms of the MZI. The spiral phase plate (SPP) placed in one of the arms of the MZI transforms the Gaussian beam into a vortex beam of order, $l_p$. The horizontal (H) polarized vortex beam and vertical (V) polarized Gaussian beam on recombination on PBS3 produces FP beam represented as $E_p = \alpha_p|E^p_H, l_p\rangle + \beta_p|E^p_V,0\rangle$ [1], where $\alpha_p$ and $\beta_p$ are the amplitude coefficient of the vortex beam and Gaussian beam, respectively, and $\alpha_p^2 + \beta_p^2 = 1$. Using the SPPs of the phase winding corresponding to vortex orders, $l_p$=1-3, we generate FP beam of order up to 3. To compensate for the pulse stretching of the ultrafast FP beam due to the dispersion of the optical components in the experiment, we have used a pair of chirped mirrors. The flip mirror directs the pump beam either to the second harmonic (SH) setup or to the projective measurement setup comprised of λ/4 plate (QWP1), λ/2 plate (HWP3), PBS4, and CCD camera for polarization and Stokes parameters. The lens L1 of focal length, $f_1$=150 mm, is used to focus the pump at the center of the dual-BIBO crystal, consists of two contiguous BIBO crystals, each having a thickness of 0.6 mm and an aperture of 1 x 1 cm$^2$ with an orthogonal optic axis [11]. Both crystals are cut at an angle, $\theta$ = 151.7° ($\varphi$ = 90°) in the optical $yz$-plane for type-I ($e+e \rightarrow o$) phase-matching for the frequency-doubling of 810 nm into 405 nm [7]. The unique geometry of the dual-BIBO crystal permits the single-pass frequency-doubling of the orthogonally polarized components of the FP beam. A lens, L2, of focal length $f_2$=100 mm, collimate the fundamental and SHG beams. Subsequently, the polarization and Stokes parameters of the SH beam extracted from the pump by the harmonic separator, S, are measured with the help of the projective measurement setup comprised of λ/4 plate (QWP2), λ/2 plate (HWP4), PBS5, and the CCD camera. The typical intensity profile of pump and SH full Poincaré beams are shown in the inset of Fig. 1.

To verify the generation of the FP beam in the pump, we have recorded the intensity distribution of the beam using different combinations of the angles of QWP1 and HWP3 and calculated the Stokes parameters, $S_1$, $S_2$, and $S_3$. Using the Stokes parameters, we have calculated the orientation ($\psi$) and ellipticity ($\chi$) [12] of the polarization ellipse with the results shown in Fig. 2. As evident from Fig. 2(a), the transverse distribution of polarization ellipse of pump beam of vortex order, $l_p$=1, contains C-point polarization singularity in the form of pair of lemon (see black square) and star (see a black circle) and a single L-line (green line) confirming the vortex order of the FP beam to be $l_p$=1. Using the experimental parameters, we have calculated the polarization distribution of the pump beam as shown in Fig. 2(b) with a pair of C-point polarization singularity and a single L-line in close agreement with the experimental result. The second column of Fig. 2, shows the magnified images of the lemon and star singularities of the experimental and theoretical polarization distribution of the FP beam with $l_p$=1. As evident from the third column, (c, d) of Fig. 2, the polarization distribution of the FP beam contains two pairs of C-point singularities and two L-lines confirming the vortex order of $l_p$=2. Using different combinations of the angles of QWP2 and HWP4, we have measured the polarization distribution of the single-pass SHG of the pump beam of $l_p$=1 and 2 with the results shown in the fourth and fifth columns of Fig. 2. As evident from the fourth column, (e, f) and fifth column, (g, h) of Fig. 2, the polarization distribution (experimental, theoretical) of the SHG beam contains two pairs of C-points, two L-lines, and four pairs of C-points, four L-lines confirming the vortex order of the SHG beam to be $l_{sh}$=2 and 4, twice that of the pump beam, respectively. Such observation confirms the doubling of the OAM mode or the C-point and L-line singularities of the pump beam in the SHG process [13]. The doubling of C-points (marked by black circle and square) and L-line (green line) singularities in the SHG process can be understood as follows. Due to the presence of the orthogonal optic axes, the dual BIBO crystal is phase-matched for the SHG of horizontal and vertical polarized beams simultaneously converting the pump FP of the electric field, $E_p=\alpha_p|E^p_H, l_p\rangle+\beta_p|E^p_V,0\rangle)$ into SHG beam electric field $E_{sh}=\alpha_{sh}|E^{sh}_V, 2l_p\rangle+\beta_{sh}|E^{sh}_H,0\rangle)$. Where $\alpha_{sh}$ and $\beta_{sh}$ are the amplitude coefficients of the second harmonic (SH) vortex and Gaussian beams, respectively, governed by the conversion efficiency of the individual beams in the nonlinear crystal. The relative phase due to the birefringence properties of the crystal is controlled by the time delay between the orthogonal components of the pump FP beam. To gain further insight into the frequency doubling characteristics of the FP beam, we have studied the Stokes phases of both pump and SHG beams. Using the FP beam of vortex order $l_p=3$, we have measured the Stokes parameters, $S_1$, $S_2$, and $S_3$ for pump and corresponding SH beam and subsequently calculated the phase distribution of Stokes fields as, $\phi_{12} = tan^{-1}(S_2/S_1)$, $\phi_{23} = tan^{-1}(S_3/S_2)$ and $\phi_{31} = tan^{-1}(S_1/S_3)$. The results are shown in Fig. 3. As evident from the first column of Fig. 3, the Stoke phase, $\phi_{12}$ representing the C-point singularity, contains three pairs of points marked by blue and red dots having phase winding corresponding to the charge $\sigma_{12} = +1$, star singularity, and -1, lemon singularity, respectively. Using the formula $\sigma_{12} = 2I_c$, the singularity indices $I_c$ of the C-points are found to be $I_c = +1/2$ and $I_c = -1/2$, respectively. The Stoke phase, $\phi_{23}$, of the FP beam, shows the azimuthal phase winding corresponding to the vortex order, $l_p = 3$ of the pump beam. Therefore, the vortex order of the FP beam can be determined from the Stoke phase, $\phi_{23}$. On the other hand, the Stoke phase, $\phi_{31}$, represents three pairs of singularities over a ring. The second column of Fig. 3 shows the Stokes phases, $\phi_{12}$, $\phi_{23}$, and $\phi_{31}$ of the SHG beam of the FP beam of vortex order, $l_p = 3$. As evident from the second column of Fig. 3, the Stokes phase, $\phi_{12}$, has six pairs (lemon, blue, and star, red) of C-point singularities with singularity indices, $I_c = +1/2$ and $I_c = -1/2$, respectively, confirming the doubling of the C-point singularities in the SHG process. On the other hand, the Stokes phase, $\phi_{23}$, shows the azimuthal phase winding corresponding to the vortex charge, $\sigma_{23} = l_{sh} = 6$, twice the order of the pump vortex charge, $l_p = 3$, of the FP beam owing to the OAM conservation in the nonlinear frequency doubling processes. Similarly, we observe six pairs of singularities over a ring in the Stoke phase, $\phi_{31}$, of the FP SHG beam. Using the experimental parameters, we have theoretically calculated the Stokes phase of the pump and corresponding SHG FP beam as shown by the third and fourth column of Fig. 3, respectively, in close agreement with the experimental results. From the results of Fig. 3, it is evident that the polarization properties of the FP beams are conserved in nonlinear processes, and the Stokes phases can be explored as important tools to characterize the nonlinear processes.

We further measured the Stokes parameters $S_1$, $S_2$, and $S_3$ of both pump and SHG beams while changing the relative intensities, $\alpha_p^2$ and $\beta_p^2 = 1-\alpha_p^2$ of the vortex and Gaussian beams, respectively, of the FP pump beam and calculate the Stoke phase, $\phi_{12}$, with the results shown in Fig. 4. As evident from the first row, (a-e), of Fig. 4, the Stoke phase, $\phi_{12}$, of the pump beam maintains a uniform phase distribution for $(\alpha_p^2, \beta_p^2) = (0,1)$ and $(1,0)$ due to the absence of orthogonal polarization states. However, the distribution of Stoke phase, $\phi_{12}$, changes with the combination of $(\alpha_p^2, \beta_p^2)$ values showing the presence of three pairs of lemon and star singularities for the $(\alpha_p^2, \beta_p^2) = (0.5, 0.5)$ and $(0.75, 0.25)$. The distribution of the Stoke phase, $\phi_{12}$, of the SHG beam, as shown in the second row, (f-j), of Fig. 4, follows the Stoke phase distribution of the pump beam with six pairs of lemon and star singularities for the $(\alpha_p^2, \beta_p^2) = (0.5, 0.5)$ and $(0.75, 0.25)$. It is evident from the first and second rows of Fig. 4 that the increase of $\alpha_p^2$ (vortex beam intensity) brings the singularity points (lemon and star) towards the center and finally annihilates each other to form the uniform Stoke phase, $\phi_{12}$ distribution. Knowing the optimum intensities of the constituent beams, we have studied the polarization coverage of the pump and corresponding SH FP beams. Using the pump FP beam of order, $l_p=3$, and $(\alpha_p^2, \beta_p^2) = (0.5, 0.5)$, we have calculated the Stokes parameters of the pump and SHG beam and projected on the surface of the Poincaré sphere with the results shown in Fig. 4(k) and Fig. 4(l), respectively. As evident from Fig. 4(k, l), the projection of Stokes parameters on the Poincaré spheres has a complicated distribution with void points. Therefore, it isn't easy to estimate the polarization coverage of the FP beam. We have simplified the problem by converting the surface of the Poincaré sphere to a 2D plane with axis orientation, ψ, and ellipticity, χ, of the polarization ellipse with the limit 0 to π, and -π/4 to +π/4, respectively. Subsequently, we divided the range of χ and ψ by the number of buckets, $n$, of equal size, counted the number of buckets having any polarization state, and divided by the total number of buckets ($n$ x $n$) to find the polarization coverage. However, the accuracy of this technique depends on the proper selection of the number of buckets. For example, suppose $n$ is considered small. In that case, the bucket size will be large enough to cover the area not having any polarization state of the beam and overestimate the polarization coverage of the beam. On the other hand, if $n$ is very large, then the bucket size will be small and fewer buckets will carry a polarization state with respect to the total number of buckets, thus underestimating the result. Therefore, to optimize the number of buckets, $n$, we locked the least count of the area measurement to be 0.1%. Such a small value of least count can be obtained by dividing the areas of the ψ-χ plane by $n$ x $n$ ~1000 buckets. Therefore, we have considered $n$~32 and measured the polarization coverage of the pump FP beam of order, $l_p = 1, 2, 3$, and 6 while varying the value of $(\alpha_p^2, \beta_p^2)$. The results are shown in Fig. 4(m). As evident from Fig. 4(m), the polarization coverage of the FP beam of all orders lies in the range of 75-95% for ($\alpha_p^2$, $\beta_p^2$) = (0.5, 0.5) and (0.75, 0.25) and around 1% for ($\alpha_p^2$, $\beta_p^2$) = (0, 1) and (1, 0). The low coverage (~1%) can be attributed to the absence of both the polarization states in the beam. On the other hand, for ($\alpha_p^2$, $\beta_p^2$) = (0.25, 0.75), the polarization coverage is varying in the range of 20-40% for all vortex orders indicating that in presence of the azimuthal phase, the relative intensity of the vortex beam plays the crucial role to the polarization coverage of the FP beams. Using the experimental parameters, we have also calculated the polarization coverage of the full Poincaré beam of different orders. As evident from Fig. 4(m), the theoretical polarization coverage (lines) is in close agreement with that of the experimental results (dots) for the full Poincaré beams of all orders. We have also measured the polarization coverage of the SH full Poincaré beam for pump orders, $l_p$ = 1, 2, and 3, for different values of ($\alpha_p^2$, $\beta_p^2$) with results shown in Fig. 4(n). As observed from Fig. 4(n), the polarization coverage of the SH FP beam follows the polarization coverage of the pump FP beam. Such results validate the direct transfer of all possible polarization states of the pump full Poincaré beam at a particular wavelength to a new wavelength through the nonlinear frequency conversion processes.

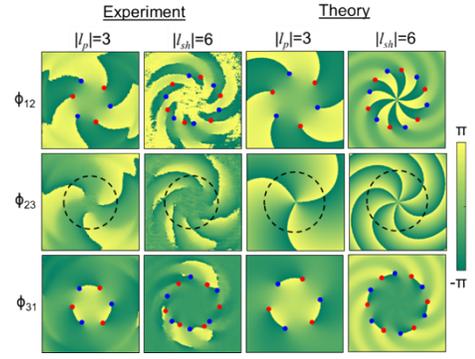

Fig. 3: Experimental Stokes phases, $\phi_{12}$, $\phi_{23}$, and $\phi_{31}$ of the pump (first column) and SHG (second column) FP beams. Theoretical Stokes phases, $\phi_{12}$, $\phi_{23}$, and $\phi_{31}$ of the pump (third column) and SHG (fourth column) FP beams. The dotted circle guides to observe phase winding.

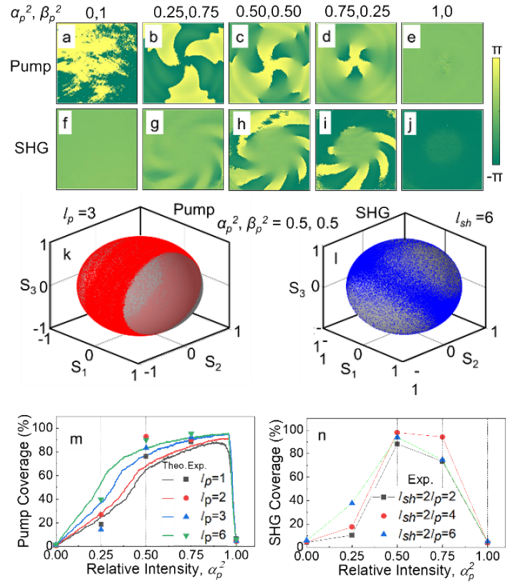

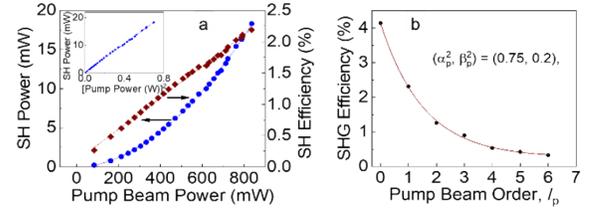

Fig. 5: (a) Variation of SH power (blue) and SH efficiency (brown) with the input power of the FP beams of vortex order, $l_p$=1. (inset) Dependence of SH power with the square of the pump power. (b) Variation of SHG efficiency with the vortex order of the FP beams. Lines are guides to the eyes.

Fig. 4: Variation of Stokes phase, $\phi_{12}$, of (a-e) pump and (f-j) SH FP beam for different combinations of ($\alpha_p^2$, $\beta_p^2$). The Stokes parameters of (k) pump and (l) SH FP beam of pump vortex order, $l_p$=3 projected on the surface of the Poincaré sphere. Variation of the polarization coverage (m) of the pump and (n) corresponding SH Poincaré beams of different pump vortex orders for a different combination of ($\alpha_p^2$, $\beta_p^2$) values. Solid lines are theoretical results.

Keeping ($\alpha_p^2$, $\beta_p^2$) = (0.5, 0.5), we have measured the power scaling of the FP beam of order, $l_p$=1, with the results shown in Figure. 5(a). As evident from Fig. 5(a), the output power and SH efficiency increase quadratic and linear, respectively, with the pump power producing a maximum average output power of 18.3 mW at a single-pass conversion efficiency as high as 2.19% without any sign of saturation. The linear variation of the SH power with the square of the pump power, as shown in the inset of Fig. 5(a), further confirms the possibility of generating an SH FP beam of higher average power with the increase of pump power. We have also measured the SHG efficiency of the FP pump beam of different orders with the results shown in Fig. 5(b). As observed in Fig. 5(b), the single-pass SHG efficiency of the FP beam decreases with the vortex order from 4.19% for $l_p$=0 (Gaussian beam with diagonal polarization) to 0.32% for $l_p$=6 similar to the single-pass SHG of the vortex beam [13] due to the increase of the dark core size with the vortex order.

In conclusion, we have studied the nonlinear generation of ultrafast FP beam at 405 nm while preserving the polarization characteristics of FP beam at the pump using two contiguous nonlinear crystals with orthogonal optic axis. Analyzing the Stokes parameters and Stokes phases, we have observed the doubling of C-points and L-lines singularities of the FP beam in the SHG process. We have devised a new method to measure the polarization coverage of the FP beam and studied the variation of the polarization coverage for different intensity weightage of the constituent beams. It is also interesting to note that the SH FP beam has the highest polarization coverage for the pump beam having equal intensity weightage of the constituent beams. To the best of our knowledge, this is the first report on the study of the effect of FP beam in the nonlinear frequency conversion process using the Stokes parameters and Stokes phases. We have also validated our experimental findings in close agreement with the theoretical study.

**Disclosures**. The authors declare no conflicts of interest.